\documentclass[aps,prd,onecolumn,groupedaddress,showpacs,nofootinbib,amssymb]{revtex4}

\usepackage[dvips]{graphicx}
\usepackage{amssymb}
\usepackage{amsmath}
\usepackage{graphicx}
\usepackage{amsfonts}
\usepackage{bm}

\newcommand{\be}{\begin{equation}}
\newcommand{\ee}{\end{equation}}
\newcommand{\bea}{\begin{eqnarray}}
\newcommand{\eea}{\end{eqnarray}}
\newcommand{\beaa}{\begin{eqnarray*}}
\newcommand{\eeaa}{\end{eqnarray*}}

\newcommand{\nn}{\nonumber \\}
\newcommand{\e}{\mathrm{e}}

\begin{document}

\title{Unimodular $F(R)$ Gravity}
\author{S.~Nojiri,$^{1,2}$\,\thanks{nojiri@gravity.phys.nagoya-u.ac.jp}
S.~D.~Odintsov,$^{3,4}$\,\thanks{odintsov@ieec.uab.es}
V.~K.~Oikonomou,$^{5,6}$\,\thanks{v.k.oikonomou1979@gmail.com}}
\affiliation{$^{1)}$ Department of Physics, Nagoya University, Nagoya 464-8602, 
Japan \\
$^{2)}$ Kobayashi-Maskawa Institute for the Origin of Particles and the
Universe, Nagoya University, Nagoya 464-8602, Japan \\
$^{3)}$Institut de Ciencies de lEspai (IEEC-CSIC),
Campus UAB, Carrer de Can Magrans, s/n\\
08193 Cerdanyola del Valles, Barcelona, Spain\\
$^{4)}$ ICREA, Passeig LluAs Companys, 23,
08010 Barcelona, Spain\\
$^{5)}$ Tomsk State Pedagogical University, 634061 Tomsk, Russia\\
$^{6)}$ Laboratory for Theoretical Cosmology, Tomsk State University of Control Systems
and Radioelectronics, 634050 Tomsk, Russia (TUSUR)\\
}

\tolerance=5000

\begin{abstract}
We extend the formalism of the Einstein-Hilbert unimodular gravity in the context of modified
$F(R)$ gravity. After appropriately modifying the Friedmann-Robertson-Walker metric in a way
that it becomes compatible to the unimodular condition of having a constant metric determinant,
we derive the equations of motion of the unimodular $F(R)$ gravity by using the metric
formalism of modified gravity with Lagrange multiplier constraint. The resulting equations are studied in frames of reconstruction method, which
enables us to realize various cosmological scenarios, which was impossible to realize in the standard
Einstein-Hilbert unimodular gravity. Several unimodular $F(R)$ inflationary
scenarios are presented, and in some cases, concordance with Planck and BICEP2 observational data can be achieved.
\end{abstract}


\maketitle

\section{Introduction}

The cosmological constant problem \cite{Peebles:2002gy} is one of the most intriguing problems in modern theoretical physics, which still remains unsolved. In principle, quantum field theoretical models predict that the cosmological constant, originating from the vacuum expectation value of certain quantum fields, is $60$-$120$ orders higher in magnitude, in comparison to the observed cosmological constant. In general relativity the cosmological constant is added by hand in the Einstein-Hilbert action, so essentially there is no intrinsic mechanism in the theory that can dynamically induce the cosmological constant. Unimodular gravity 
\cite{Anderson:1971pn,Buchmuller:1988wx,Henneaux:1989zc,Unruh:1988in,Ng:1990xz,Finkelstein:2000pg,Alvarez:2005iy,Alvarez:2006uu,
Abbassi:2007bq,Ellis:2010uc,Jain:2012cw,Singh:2012sx,Kluson:2014esa,Padilla:2014yea,Barcelo:2014mua,Barcelo:2014qva,
Burger:2015kie,Alvarez:2015sba,Jain:2012gc,Jain:2011jc,Cho:2014taa,Basak:2015swx,Gao:2014nia,Eichhorn:2015bna,Saltas:2014cta} offers an interesting and conceptually simple proposal for the cosmological constant problem, since it is a particular case of general relativity, so no new physical assumptions are required. However, it can be suggested that some strong claims which state that the cosmological constant is solved by using the unimodular gravity formalism, should be weakened to some extent, like for example in Ref. \cite{Padilla:2014yea}. It would be more appropriate to state that unimodular gravity historically offered a proposal to solve the cosmological constant problem, but this cannot be considered to be the full solution of the cosmological constant problem.

In unimodular gravity, the cosmological constant originates from the trace-free part if the Einstein field equations, so the cosmological constant appears in the theory without adding it by hand, and more importantly it is not related to the vacuum expectation value of some matter field. Therefore, in principle the cosmological constant can be fixed from the observed value, hence providing a somewhat artificial solution to the cosmological constant problem. In unimodular gravity, a trace free part of the Einstein equations can be obtained, if the determinant of the metric $\sqrt{-g}$ is some fixed number, or more generally a function. Also in the context of unimodular gravity the problem of late-time acceleration can be addressed, in which case the full metric is decomposed in two parts, the unimodular metric and a scalar field \cite{Jain:2012gc,Jain:2011jc,Cho:2014taa}. Also, in the context of unimodular gravity, the cosmological perturbations of the comoving curvature perturbation originating from primordial quantum fluctuations, are the same as in ordinary general relativity, at least when linear perturbation theory is used \cite{Basak:2015swx,Gao:2014nia}. However in some cases, a difference in the perturbations can be found, if certain gauges choices are made, due to the constraint of the metric determinant \cite{Basak:2015swx}. Also, in Ref.~\cite{Gao:2014nia}, a thorough analysis of the primordial curvature perturbations in the context of unimodular gravity revealed that no strong constraints on the theory can be obtained. Particularly, for adiabatic fluctuations, the differences between unimodular gravity and the standard general relativity are suppressed on large angular scales. However, some differences can be traced on the Sachs-Wolfe relation between the gravitational potential and the microwave temperature anisotropies, due to the fact that the shift variable cannot be set equal to zero in the unimodular gravity case \cite{Gao:2014nia}. 

Owing to the qualitatively interesting implications of the unimodular gravity approach, in this paper we shall consider the $F(R)$ gravity extension of Einstein-Hilbert unimodular gravity. So we shall use the basic assumption of the unimodular gravity, namely the constrained determinant of the metric, and we shall extend the Jordan frame $F(R)$ gravity formalism by taking into account this constraint. The $F(R)$ gravity theories (for reviews see \cite{Nojiri:2010wj,Nojiri:2013zza,Nojiri:2006ri,Capozziello:2010zz,Capozziello:2011et,Bamba:2012cp}) have quite appealing features, such as predicting late-time acceleration and early-time acceleration in a unified way \cite{Nojiri:2003ft}, but also many exotic cosmologies for the standard Einstein-Hilbert general relativity, can be realized in $F(R)$ gravity theories, such as bouncing cosmologies \cite{Novello:2008ra,Li:2014era,Brandenberger:1993ef,Mukhanov:1991zn,Cai:2013kja,Cai:2014zga,deHaro:2014kxa,Qiu:2010ch,Qiu:2010vk,Lehners:2011kr,Khoury:2012dn,Koehn:2013upa,Easson:2011zy,Cai:2012va,Odintsov:2015zua,Lehners:2015mra,Cai:2007qw}, see for example 
\cite{Odintsov:2014gea,Bamba:2013fha,Barragan:2009sq,Oikonomou:2014jua,Odintsov:2015uca,Elizalde:2014uba,Escofet:2015gpa,Oikonomou:2015qha,Odintsov:2015zua} for some recent works on these issues. Also, for an important stream of research papers on $F(R)$ gravity, see 
Refs.~\cite{Dunsby:2010wg,Capozziello:2003gx,Hu:2007nk,Capozziello:2005ku,Faraoni:2006sy,Olmo:2005zr,Appleby:2007vb} and references therein. Our purpose in this paper is to investigate all the new features that the unimodular constraint on the metric imposes on the $F(R)$ gravity theories. As we demonstrate, the flat Friedmann-Robertson-walker metric is not compatible with the unimodular constraint, so we appropriately fix the metric, in order to satisfy the unimodular constraint. Then, we derive the gravitational equations of motion by using the metric $F(R)$ gravity formalism, and by treating the unimodular constraint by introducing a Lagrange multiplier. The unimodular $F(R)$ gravity formalism results to a new reconstruction method which provides us with two appealing features: First it is possible to realize any cosmological evolution with a specific Hubble rate, and investigate which unimodular $F(R)$ can yield such a cosmological evolution, and secondly, it is possible to fix the $F(R)$ gravity and investigate which cosmological evolution is realized from this $F(R)$ gravity. As we shall demonstrate, by using some characteristic examples, the resulting picture is different in comparison to both standard general relativity unimodular gravity and to $F(R)$ gravity without the unimodular constraint. We also investigate how some inflation-related cosmologies \cite{Mukhanov:1990me,Gorbunov:2011zzc,Linde:2014nna,Bamba:2015uma,Lyth:1998xn,Miao:2015oba} can be realized in the context of unimodular $F(R)$ gravity and we provide a rough estimate of the observational indices. Particularly, we calculate the Hubble slow-roll parameters 
\cite{Liddle:1994dx,Liddle:1992wi,Copeland:1993jj} and we compute the spectral index of primordial curvature perturbations \cite{Mukhanov:1990me,Brandenberger:1983tg,Brandenberger:1983vj,Baumann:2009ds} and the scalar-to-tensor ratio related to the Hubble slow-roll indices. Also, by considering the $F(R)$ gravity in perfect fluid description, we calculate the slow-roll indices and the corresponding observational indices in the context of a perfect fluid $F(R)$ gravity and as we demonstrate, concordance with the recent Planck \cite{Ade:2015lrj,Planck:2013jfk} and BICEP2 \cite{Array:2015xqh} data can be achieved in some cases. The most important outcome of this paper is the new reconstruction method, which enables us to realize various cosmological scenarios which was impossible to realize in the context of Einstein-Hilbert gravity. For other existing reconstruction methods in the context of $F(R)$ gravities, see 
\cite{Nojiri:2010wj,Nojiri:2013zza,Nojiri:2006ri,Capozziello:2010zz,Capozziello:2011et,Bamba:2012cp,Nojiri:2006gh,Nojiri:2006be,
Capozziello:2006dj,Bamba:2008ut}.

This paper is organized as follows: In section II A, we discuss the general implications of the unimodular constraint on a Friedmann-Robertson-Walker metric, and we investigate how the metric should be fixed in order to satisfy the unimodular constraint. In section II B, we present in detail the unimodular $F(R)$ gravity formalism and we discuss the implications of the unimodular constraint on the metric $F(R)$ formalism. In section II C, we present some concrete examples of cosmologies that can be realized in the context of unimodular $F(R)$ gravity, and we investigate which unimodular $F(R)$ gravities can realize these cosmologies. We also present some examples of inflationary cosmologies and we provide a rough estimate of the observational indices. In section II D, we treat the unimodular $F(R)$ gravity as a perfect fluid and we calculate the slow-roll indices and the corresponding observational indices. As we demonstrate, in some cases, concordance with current observational data is achieved. Finally, the conclusions follow in the end of the paper.

\section{The $F(R)$ Gravity Description of Unimodular Gravity}

\subsection{General Considerations}

In this section we shall generalize the unimodular Einstein-Hilbert gravity formalism of 
Refs.~\cite{Anderson:1971pn,Buchmuller:1988wx,Henneaux:1989zc,Unruh:1988in,
Ng:1990xz,Finkelstein:2000pg,Alvarez:2005iy,Alvarez:2006uu,Abbassi:2007bq,Ellis:2010uc,
Jain:2012cw,Singh:2012sx,Kluson:2014esa,Padilla:2014yea,Barcelo:2014mua,Barcelo:2014qva,
Burger:2015kie,Alvarez:2015sba,Jain:2012gc,Jain:2011jc,Cho:2014taa,Basak:2015swx,Gao:2014nia}, in order to provide the $F(R)$ gravity formalism of unimodular gravity. As in the standard Einstein-Hilbert unimodular gravity, the unimodular gravity approach is based on the assumption that the determinant of the metric tensor is fixed, so that $g_{\mu \nu}\delta g^{\mu \nu}=0$, which means that the components of the metric are adjusted in such a way so that $\sqrt{-g}$ is fixed. Throughout this paper we shall assume that the determinant of the metric is constrained to obey the following relation, 
\be
\label{Uni1}
\sqrt{-g}=1\, .
\ee
Usually in most cosmological studies, for simplicity and also because the Universe seems to be nearly flat, the metric is assumed to be a flat Friedman-Robertson-Walker (FRW) metric with line element,
\be
\label{FRW}
ds^2 = - dt^2 + a(t)^2 \sum_{i=1}^3 \left( dx^i \right)^2 \, .
\ee
However, as it can be easily checked, the unimodular constraint of Eq. (\ref{Uni1}) is not satisfied by the metric (\ref{FRW}). So the FRW metric has to be redefined in such a way so that the unimodular constraint (\ref{Uni1}) is satisfied, so we redefine the cosmic time coordinate in the following way,
\be
\label{Uni2}
d\tau = a(t)^3 dt\, .
\ee
Correspondingly, the FRW metric (\ref{FRW}) can be rewritten in the following way,
\be
\label{UniFRW}
ds^2 = - a\left(t\left(\tau\right)\right)^{-6} d\tau^2 + a\left(t\left(\tau\right)\right)^{2} 
\sum_{i=1}^3 \left( dx^i \right)^2 \, ,
\ee
and it can be easily checked, for the metric (\ref{UniFRW}), the unimodular constraint (\ref{Uni1}) is satisfied. In the following we shall refer to the metric of Eq.~(\ref{UniFRW}) as the unimodular FRW metric. Let us for example consider the case that the FRW metric describes a de Sitter expanding Universe, for which the scale factor is, 
\be
\label{Uni3}
a(t) = \e^{H_0 t}\, ,
\ee
with $H_0$ being an arbitrary constant. In this case, by using Eq.~(\ref{Uni2}), we obtain the relation between the variable $\tau$ and the cosmic time, which is, 
\be
\label{Uni4}
\tau = \frac{1}{3H_0} \e^{3H_0 t}\, , 
\ee
so by substituting in Eq.~(\ref{Uni3}), we obtain the scale factor of the Universe in terms of the variable $\tau$, which reads,
\begin{equation}\label{scalenew1}
a(\tau) \equiv a\left(t\left(\tau\right)\right) = \left( 3H_0 \tau \right)^\frac{1}{3}\, ,
\end{equation}
and therefore, by substituting in the unimodular FRW metric (\ref{UniFRW}), the latter takes the following form,
\be
\label{Uni5}
ds^2 = - \left( 3 H_0 \tau \right)^{-2} d\tau^2 
+ \left( 3 H_0 \tau \right)^\frac{2}{3} \sum_{i=1}^3 \left( dx^i \right)^2 \, .
\ee
Consider another example, for which the Universe is described by a power-law scale factor of the form,
\be
\label{Uni6}
a(t) = \left( \frac{t}{t_0} \right)^{h_0}\, ,
\ee
where $t_0$ and $h_0$ are constants. In this case, by using Eq.~(\ref{Uni2}), we get,
\be
\label{Uni7}
\tau = \frac{t_0}{3 h_0 + 1} \left( \frac{t}{t_0} \right)^{3 h_0 + 1}\, ,
\ee
and by substituting in Eq.~(\ref{Uni6}), we easily get the scale factor in terms of the variable $\tau$,
\begin{equation}\label{scalenew2}
a(\tau) = \left( \frac{\left( 3 h_0 + 1 \right) \tau}{t_0} \right)^\frac{h_0}{3h_0 + 1}\, .
\end{equation} 
Therefore, the corresponding unimodular FRW metric (\ref{UniFRW}) takes the following form,
\be
\label{Uni8}
ds^2 = - \left( \frac{\left( 3 h_0 + 1 \right) \tau}{t_0} \right)^\frac{-6h_0}{3h_0 + 1}d\tau^2 
+ \left( \frac{\left( 3 h_0 + 1 \right) \tau}{t_0} \right)^\frac{2h_0}{3h_0 + 1} 
\sum_{i=1}^3 \left( dx^i \right)^2 \, .
\ee
We should note that the metric (\ref{Uni5}) becomes identical to the de Sitter space metric, if the constants are assumed to satisfy $h_0, t_0 \to \infty$, by also keeping $\frac{\left( 3 h_0 + 1 \right) \tau}{t_0}$ to be finite, so that the following limiting case is achieved, 
\be
\label{Uni9}
\frac{\left( 3 h_0 + 1 \right) \tau}{t_0} \to 3 H_0\, .
\ee
Then by using the unified geometric framework implied by the metrics (\ref{Uni5}) and (\ref{Uni8}), we construct the following metric, 
\be
\label{Uni10}
ds^2 = - \left( \frac{\tau}{\tau_0} \right)^{-6f_0}d\tau^2 
+ \left( \frac{\tau}{\tau_0} \right)^{2f_0} 
\sum_{i=1}^3 \left( dx^i \right)^2 \, ,
\ee
with the parameters $\tau_0$ and $f_0$ being arbitrary real constant numbers. So in the case that $f_0=\frac{1}{3}$, then this corresponds to a de Sitter cosmological evolution. Correspondingly, if $\frac{1}{4}\leq f_0 < \frac{1}{3}$, which occurs when, $h_0\geq 1$, the Universe is described by a quintessence evolution. Moreover, in the case $0<f_0<\frac{1}{4}$, which implies, $0<h_0<1$, the Universe is expanding in a decelerating way. Finally, when $f_0<0$ or if $f_0>\frac{1}{3}$, which implies, $h_0<0$, the Universe is described by a phantom evolution \cite{Bamba:2008hq,Caldwell:2003vq,Elizalde:2004mq,Liu:2012iba,Piao:2004tq,Guo:2004ae,Nesseris:2006er,Chen:2008ft}.

\subsection{Unimodular $F(R)$ Gravity Action and a General Reconstruction Method}

Having discussed the essential geometric conventions of unimodular gravity, with regards to the spacetime metric, in this section we proceed to present the $F(R)$ gravity generalization of unimodular gravity. Our aim is to present the general unimodular $F(R)$ gravity function in the Jordan frame, deferring the Einstein frame study to a future work. As we demonstrate, the unimodular $F(R)$ gravity formalism provides us with a quite general reconstruction technique, which enables us to realize a large number of cosmological evolution scenarios, mainly those for which 
Eq.~(\ref{Uni2}) can be explicitly solved with respect to the cosmic time $t$. We assume that the metric of the Universe is the one appearing in 
Eq.~(\ref{UniFRW}), and in the following, we denote the metric tensor as $g_{\mu \nu}$ and its determinant by $g$. We shall employ the Lagrange multiplier method \cite{Lim:2010yk,Capozziello:2013xn,Capozziello:2010uv} in order to satisfy the unimodular constraint, as we now evince. The Jordan frame unimodular $F(R)$ gravity function is\footnote{It is expected that unimodular gravity is realized mainly as a quantum gravity. Hence,
one may conjecture that constraint in Eq. (\ref{Uni11}) has a more complicated time-dependent form and quickly decays as the Universe expands, so that gravity takes it'w well known form.}, 
\be
\label{Uni11}
S = \int d^4 x \left\{ \sqrt{-g} \left( F(R) - \lambda \right) + \lambda \right\} 
+ S_\mathrm{matter} \, ,
\ee
where $F(R)$ is a suitable differentiable function of the Ricci scalar curvature $R$, and $\lambda$ is the Lagrange multiplier function, which when the action is varied with respect to it, the variation yields the unimodular constraint (\ref{Uni1}). Moreover, the term $S_\mathrm{matter}$ contains all the matter fluids present, which are assumed to be perfect fluids. Varying the action (\ref{Uni11}) with respect to the metric, we obtain the following equation of motion,
\be
\label{Uni12}
0=\frac{1}{2}g_{\mu\nu} \left( F(R) - \lambda \right) - R_{\mu\nu} F'(R) 
+ \nabla_\mu \nabla_\nu F'(R) - g_{\mu\nu}\nabla^2 F'(R) + \frac{1}{2} T_{\mu\nu} \, ,
\ee
where $T_{\mu\nu}$ stands for the energy-momentum tensor of the matter fluids. For the unimodular metric of Eq.~(\ref{UniFRW}), the non-vanishing components of the Levi-Civita connection and of the curvatures are given below,
\begin{align}
\label{Uni13}
& \Gamma^t_{tt} = - 3 K\, , \quad \Gamma^t_{ij} = a^8 K \delta_{ij}\, , \quad 
\Gamma^i_{jt} = \Gamma^i_{tj} = K \delta_j^{\ i} \, , \nn
& R_{tt} - 3 \dot K - 12 K^2\, , \quad 
R_{ij} = a^8 \left( \dot K + 6 K^2 \right) \delta_{ij}\, ,
\end{align}
where we introduced the function $K(\tau)$, which is equal to $K\equiv \frac{1}{a} \frac{da}{d\tau}$, so it is the analog Hubble rate for the $\tau$ variable. In addition, the corresponding Ricci scalar curvature corresponding to the metric (\ref{UniFRW}), is equal to,
\begin{equation}\label{scalarunifrw}
R = a^6 \left( 6 \dot K + 30 K^2 \right) \, ,
\end{equation}
which we extensively use in the following sections. By taking into account Eq.~(\ref{Uni13}), the $(t,i)$ and $(i,t)$ components of the equations of motion (\ref{Uni12}), identically vanish, and furthermore, the $(t,t)$ and $(i,j)$ components yield the following equations, 
\begin{align}
\label{Uni14}
0 = & - \frac{a^{-6}}{2} \left( F(R) - \lambda \right) + \left( 3 \dot K + 12 K^2 \right) F'(R) 
 - 3 K \frac{d F'(R)}{d\tau} + \frac{a^{-6}}{2} \rho \, , \\
\label{Uni15}
0 = & \frac{a^{-6}}{2} \left( F(R) - \lambda \right) - \left( \dot K + 6 K^2 \right) F'(R) 
+ 5 K \frac{d F'(R)}{d\tau} + \frac{d^2 F' (R)}{d\tau^2} + \frac{a^{-6}}{2} p \, .
\end{align}
where we assumed that the energy momentum tensor is described by a perfect fluid with energy density $\rho$ and pressure $p$. Moreover, the ``prime'' in the above equations denotes differentiation with respect to the Ricci scalar, while the ``dot'' denotes differentiation with respect to the time coordinate $\tau$. By combining Eqs.~(\ref{Uni14}) and (\ref{Uni15}),  we obtain the following differential equation,
\be
\label{Uni16}
0 = \left( 2 \dot K + 6 K^2 \right) F'(R) 
+ 2 K \frac{d F'(R)}{d\tau} + \frac{d^2 F' (R)}{d\tau^2} + \frac{a^{-6}}{2} \left( \rho + p \right) \, .
\ee
The above differential equation plays a central role in the unimodular reconstruction technique, since it enables us to realize a large number of cosmological scenarios. Actually, by specifying the scale factor $a(\tau)$ of a specific cosmological evolution, then by inserting the scale factor in the differential equation (\ref{Uni16}), we can solve it and find the $F(R)$ gravity that generates this evolution. Conversely, for a given $F(R)$ gravity, we can find which cosmological evolution it can generate, by solving the differential equation (\ref{Uni16}). In addition to this equation, a central role in the unimodular $F(R)$ gravity reconstruction technique is played by both equations (\ref{Uni14}) or (\ref{Uni15}), which separately yield the Lagrange multiplier $\lambda$. It is worth describing the reconstruction technique in more detail, so for example if the cosmological evolution is specified to have the scale factor $a=a(\tau)$, by plugging the scale factor in the differential equation (\ref{Uni16}), we can solve it and obtain the $F'(R(\tau))$ function which realizes this cosmological evolution, since the differential equation (\ref{Uni16}) can be regarded as a differential equation which actually yields the function $F'(R)$ as a function of $\tau$. The resulting solution will be of the form $F' = F'(\tau)$, so by inverting the Ricci scalar function $R=R(\tau)$ of Eq.~(\ref{scalarunifrw}), we may obtain the function $\tau=\tau (R)$. Consequently, by substituting the function $\tau=\tau (R)$ in the resulting $F'(\tau)$ function, we finally obtain the function $F'(R)=F'\left( \tau\left(R\right) \right)$, which when integrated with respect to the Ricci scalar, it yields the $F(R)$ gravity which realizes the cosmological evolution with scale factor $a=a(\tau)$.

\subsection{Applications of the Reconstruction Method: Realization of General Cosmological Evolutions}

Having described all the essential features of the unimodular $F(R)$ gravity and the corresponding reconstruction method, let us now demonstrate how various cosmological scenarios can be realized in the context of unimodular $F(R)$ gravity. We start off with the cosmological evolution described by the metric (\ref{Uni10}), which in view of Eq.~(\ref{UniFRW}) it implies that,
\be
\label{Uni17}
a(\tau) = \left( \frac{\tau}{\tau_0} \right)^{f_0}\, , \quad K = \frac{f_0}{\tau} \, ,
\ee
and in addition the corresponding Ricci scalar curvature is equal to,
\be
\label{Uni18}
R = \frac{ - 6f_0 + 30 f_0^2 }{\tau_0^2} \left( \frac{\tau}{\tau_0} \right)^{6f_0 - 2}\, .
\ee
Correspondingly, the differential equation given in Eq.~(\ref{Uni16}), takes the following form,
\be
\label{Uni19}
0= \frac{ - 2 f_0 + 6 h_0^2 }{\tau^2} F'(\tau) + \frac{2f_0}{\tau} \frac{d F'(\tau)}{d\tau}  
+ \frac{d^2 F'(\tau)}{d\tau^2} + \frac{1}{2} \left( \frac{\tau}{\tau_0} \right)^{-6f_0}
\left( \rho + p \right) \, .
\ee
Let us first consider how the cosmological evolution (\ref{Uni17}) can be realized from the vacuum unimodular $F(R)$ gravity, in which case the contribution coming from perfect matter fluids can be neglected, that is, $\rho=p=0$. In this case, the general solution of differential equation (\ref{Uni19}) is equal to,
\be
\label{Uni20}
F'(\tau) \propto \mathcal{C}_+\left( \frac{\tau}{\tau_0} \right)^{\alpha_+}+\mathcal{C}_- \left( \frac{\tau}{\tau_0} \right)^{\alpha_-}\, , \quad 
\alpha_\pm \frac{ - 2 f_0 + 1 \pm \sqrt{ - 20 f_0^2 + 4 f_0 + 1 }}{2} \, ,
\ee
where the parameters $\mathcal{C}_+$ and $\mathcal{C}_-$ are arbitrary integration constants. By using the expression for the Ricci scalar (\ref{scalarcurvatau}), and inverting the function $R=R(\tau)$, the expression in Eq.~(\ref{Uni20}) can be written as follows, 
\be
\label{Uni21}
F'(R) =\mathcal{C}_+\left(\frac{\tau_0}{30f_0^2-6f_0}\right)^{\frac{\alpha_+}{6f_0-2}} R^{\frac{\alpha_+}{6 f_0 - 2}}+\mathcal{C}_-\left(\frac{\tau_0}{30f_0^2-6f_0}\right)^{\frac{\alpha_-}{6f_0-2}} R^{\frac{\alpha_-}{6 f_0 - 2}}\, .
\ee
so eventually, by integrating once with respect to the Ricci scalar, we obtain the $F(R)$ gravity, which is
\begin{equation}\label{rfrfgravfinal}
F(R) =\mathcal{C}_+\frac{6f_0-2}{a_++6f_0-2}\left(\frac{\tau_0}{30f_0^2-6f_0}\right)^{\frac{\alpha_+}{6f_0-2}} R^{\frac{\alpha_+}{6 f_0 - 2}+1}+\mathcal{C}_-\frac{6f_0-2}{a_-+6f_0-2}\left(\frac{\tau_0}{30f_0^2-6f_0}\right)^{\frac{\alpha_-}{6f_0-2}} R^{\frac{\alpha_-}{6 f_0 - 2}+1}\, .
\end{equation}
By introducing for notational simplicity the parameters $\mathcal{A}_{\pm}$, which are defined to be,
\begin{equation}\label{aplusmineus}
\mathcal{A_{\pm}}=\mathcal{C}_{\pm}\frac{6f_0-2}{a_{\pm}+6f_0-2}\left(\frac{\tau_0}{30f_0^2-6f_0}\right)^{\frac{\alpha_{\pm}}{6f_0-2}} \, ,
\end{equation}
by using Eq.~(\ref{Uni14}), the unimodular Lagrange multiplier function $\lambda$ reads,
\begin{align}
\label{lambdafunctionl}
\lambda (\tau )= & \left(6^{\frac{\alpha_-}{-2+8 f_0}} \mathcal{A}_- \left(\frac{f_0 (-1+5 f_0) \left(\frac{\tau }{\tau_0}\right)^{8 f_0}}{\tau ^2}\right)^{\frac{\alpha_-}{-2+8 f_0}}+6^{\frac{\alpha_+}{-2+8 f_0}} \mathcal{A}_+ \left(\frac{f_0 (-1+5 f_0) \left(\frac{\tau }{\tau_0}\right)^{8 f_0}}{\tau ^2}\right)^{\frac{\alpha_+}{-2+8 f_0}}\right) \nn 
& \times \left(-\tau ^2+6 f_0 (-1+4 f_0-\tau ) \left(\frac{\tau }{\tau_0}\right)^{6 f_0}\right) \left(\frac{\tau }{\tau_0}\right)^{8 f_0}\tau ^{-4}6 f_0 (-1+5 f_0) \, . 
\end{align}
Note that in the case in which the parameter $f_0$ satisfies the inequality $- 20 f_0^2 + 4 f_0 + 1<0$, the parameters $\alpha_\pm$ become complex numbers. In this case, by separating $\alpha_\pm$ into real and an imaginary parts,  that is, $\alpha_\pm = \alpha_R \pm i \alpha_I$, the resulting $F'(R)$ unimodular gravity is given by,
\be
\label{Uni22} 
F'(R) =R^{\frac{\alpha_R}{6 f_0 - 2}} \left( C R^{i\frac{\alpha_I}{6 f_0 - 2}}
+ C^* R^{-i\frac{\alpha_I}{6 f_0 - 2}} \right) \, .
\ee
Here $C$ is a complex number and $C^*$ is the complex conjugate of $C$ so that $F'(R)$ is a real function of $R$. 

We proceed to investigate which unimodular $F(R)$ gravity generates the cosmological evolution (\ref{Uni17}), in the case that a perfect matter fluid is present, with equation of state $p=w\rho$, where $w$ is the constant equation of state parameter. Then by using the energy momentum tensor conservation law $\nabla_\nu T^{\mu \nu} =0$, we find that the energy density is expressed as a function of the time parameter $\tau$ as follows,
\be
\label{Uni23}
\rho = \rho_0 a^{- 3 \left( w + 1 \right)} 
= \left( \frac{\tau}{\tau_0} \right)^{- 3 \left( w + 1 \right)f_0}\, .
\ee
and therefore, the general solution of the differential equation (\ref{Uni19}) is given by, 
\begin{align}
\label{Uni24}
F'(R) =& f_0 \left( \frac{\tau_0^2 R}{ - 6 f_0 + 30 f_0^2 } 
\right)^{\frac{2 - \left( 3 w + 9 \right) f_0}{6 f_0 - 2}} 
+ C_+  R^{\frac{\alpha_+}{6 f_0 - 2}}
+ C_-  R^{\frac{\alpha_-}{6 f_0 - 2}} \, , \nn
f_0 = & - \frac{1+w}{2} \frac{\rho_0 \tau_0^2}
{  69 f_0^2 - 25 f_0 + 2 + \left( 51 - 9 f_0 \right) w + 9 f_0^2 w^2 } \, ,
\end{align}
where the parameters $C_\pm$ are arbitrary integration constants. 

As another example, we may consider the following variation from the model of Eq.~(\ref{Uni17}), in which case the scale factor is assumed to be, 
\be
\label{Uni25}
a(\tau) = \left( \frac{\tau}{\tau_0} \right)^{f_0} \left( 1 + b(\tau) \right) \, , 
\left| b(\tau) \right| \ll 1 \, ,
\ee
which may correspond to the slow-roll inflation case, with $f_0=\frac{1}{3}$, that is, the de Sitter 
space-time which corresponds to the $b(\tau)=0$ case. For this case we are mainly interested in the inflationary era, so we neglect the contribution from any matter fluids, so we assume $\rho = p = 0$. We now introduce the variation from the solution appearing in Eq.~(\ref{Uni20}) which is contained in the term $f(\tau)$, as follows:
\be
\label{Uni26}
F'(\tau) = A_+ \left( \frac{\tau}{\tau_0} \right)^{\alpha_+}
+ A_- \left( \frac{\tau}{\tau_0} \right)^{\alpha_-} + f(\tau) \, .
\ee
Then Eq.~(\ref{Uni16}) takes the following form, 
\begin{align}
\label{Uni27}
0= & \frac{ - 2 f_0 + 6 f_0^2 }{\tau^2} f(\tau) + \frac{2f_0}{\tau} \frac{d f(\tau)}{d\tau} 
+ \frac{d^2 f(\tau)}{d\tau^2} - \gamma(\tau) \, , \nn
\gamma (\tau) \equiv & - \left( \frac{d^2 b(\tau)}{d\tau^2} + \frac{2f_0}{\tau} 
\frac{db(\tau)}{d\tau} \right) \left( A_+ \left( \frac{\tau}{\tau_0} \right)^{\alpha_+}
+ A_- \left( \frac{\tau}{\tau_0} \right)^{\alpha_-} \right) 
 \\ \notag &- \frac{1}{\tau} \frac{df(\tau)}{d\tau} \left( 
\alpha_+ A_+ \left( \frac{\tau}{\tau_0} \right)^{\alpha_+}
+ \alpha_- A_- \left( \frac{\tau}{\tau_0} \right)^{\alpha_-} \right. \nn
& \left. + \frac{1}{\tau^2}\left( \alpha_+ \left( \alpha_+ - 1 \right)
A_+ \left( \frac{\tau}{\tau_0} \right)^{\alpha_+}
+ \alpha_- \left( \alpha_- - 1 \right) A_- \left( \frac{\tau}{\tau_0} \right)^{\alpha_-}
\right)\right) + \mathcal{O} \left( b(\tau)^2 \right)\, .
\end{align}
Consequently, the general solution of Eq.~(\ref{Uni27}) is equal to,
\be
\label{Uni28}
f(\tau) = \left( \frac{\tau}{\tau_0} \right)^{\alpha_+} \int^\tau d\tau' 
\frac{\tau_0}{\alpha_- - \alpha_+} \left( \frac{\tau'}{\tau_0} \right)^{1- \alpha_+}
\gamma ( \tau' ) 
 - \left( \frac{\tau}{\tau_0} \right)^{\alpha_-} \int^\tau d\tau' 
\frac{\tau_0}{\alpha_- - \alpha_+} \left( \frac{\tau'}{\tau_0} \right)^{1- \alpha_-}
\gamma ( \tau' ) \, .
\ee
On the other hand, by using Eq.~(\ref{Uni25}), we find the following expression for the 
scalar Ricci curvature $R$,
\be
\label{Uni29}
R = \frac{ - 6 f_0 + 30 f_0^2 }{\tau_0^2} \left( \frac{\tau}{\tau_0} \right)^{6f_0 - 2} 
+ \left( \frac{\tau}{\tau_0} \right)^{6f_0} \left\{ 
\frac{6 \left(- 6 f_0 + 30 f_0^2 \right)}{\tau^2} b(\tau) 
+ \frac{60 f_0}{\tau} \frac{d b(\tau)}{d\tau} + 6 \frac{d^2 b(\tau)}{d\tau^2} \right\} 
+ \mathcal{O}\left( b(\tau)^2 \right)\, ,
\ee
which can be solved with respect to $\tau$ to yield the function $\tau=\tau (R)$, which is, 
\begin{align}
\label{Uni30}
\tau =& \tau_0 \left( \frac{\tau_0^2 R}{ - 6 f_0 + 30 f_0^2 } 
\right)^{\frac{1}{6 f_0 - 2}}\times \\ \notag &
\left( 1 - \frac{1}{\left( 6 f_0 - 2 \right) R} 
\left( \frac{\tau_0^2 R}{ - 6 f_0 + 30 f_0^2 } \right)^{- \frac{6f_0}{6 f_0 - 2}}
\left\{ \frac{6 \left( - 6 f_0 + 30 f_0^2 \right)}{\tau_0^2} \left( \frac{\tau_0^2 R}{ - 6 f_0 
+ 30 f_0^2 } \right)^{- \frac{2}{6 f_0 - 2}} \right. \right. \nn
& \times b \left( 
\tau_0 \left( \frac{\tau_0^2 R}{ - 6 f_0 + 30 f_0^2 } \right)^{\frac{1}{6 f_0 - 2}} \right) 
+ \frac{60 f_0}{\tau_0} \left( \frac{\tau_0^2 R}{ - 6 f_0 
+ 30 f_0^2 } \right)^{-\frac{1}{6 f_0 - 2}} b' \left( \tau_0 \left( 
\frac{\tau_0^2 R}{ - 6 f_0 + 30 f_0^2 } \right)^{\frac{1}{6 f_0 - 2}} \right) \nn
& \left. \left. + 6 b'' \left( \tau_0 
\left( \frac{\tau_0^2 R}{ - 6 f_0 + 30 f_0^2 } \right)^{\frac{1}{6 f_0 - 2}} \right) \right\} 
\right) + \mathcal{O}\left( b(\tau)^2 \right) \, ,
\end{align}
where we assumed that $f_0 \neq \frac{1}{3}$. For the case $f_0 \neq \frac{1}{3}$, which corresponds to the quasi-de Sitter 
spacetime, the Eq.~(\ref{Uni29}) reduces to the following,
\be
\label{Uni31}
R = \frac{ 8 }{3\tau_0^2} + \left( \frac{\tau}{\tau_0} \right)^2 \left\{ 
\frac{16 }{\tau^2} b(\tau) 
+ \frac{20}{\tau} \frac{d b(\tau)}{d\tau} + 6 \frac{d^2 b(\tau)}{d\tau^2} \right\} 
+ \mathcal{O}\left( b(\tau)^2 \right)\, ,
\ee
which can be solved with respect to $\tau$, and then the function $\tau=\tau (R)$ is obtained. By substituting the resulting expression of the function $\tau (R)$ in Eq.~(\ref{Uni26}), and also taking into account Eq.~(\ref{Uni28}), we finally obtain the resulting expression for the $F'(R)$ gravity, and eventually the $F(R)$ gravity. 

Let us present another example with relatively interesting phenomenology, with it's Hubble rate as a function of the cosmic time being,
\begin{equation}\label{eq}
H(t)=H_0-\frac{M^2}{6}(t-t_i)\, ,
\end{equation}
with $H_0$, $M$, and $t_i$ being arbitrary constants. The model (\ref{eq}) describes the $R^2$ inflation Starobinsky model \cite{Starobinsky:1980te,Barrow:1988xh,Odintsov:2015zza,Odintsov:2015gba}, and the corresponding scale factor is,
\begin{equation}\label{scalestaron}
a(t)=a_0 \e^{H_0(t-t_i)-\frac{M^2}{12}(t-t_i)^2}\, .
\end{equation}
Since the study of this model is done for early times, we can approximate the scale factor (\ref{scalestaron}) as follows,
\begin{equation}\label{scalestaron1}
a(t)=a_0 \e^{H_0(t-t_i)+\frac{M^2}{6}tt_i}\, .
\end{equation}
which easily follows since $t_i,t\ll \frac{1}{H_0}$. 
Then, by using Eq.~(\ref{Uni2}), we easily obtain that the coordinate $\tau$ is related to the cosmic time $t$, as follows,
\begin{equation}\label{tauoft}
\tau (t)=\frac{2 a_0^3 \e^{-3 H_0 t_i+\frac{1}{2} t \left(6 H_0+M^2 t_i\right)}}{6 H_0+M^2 t_i}\, ,
\end{equation}  
which can be easily inverted to yield the function $t=t(\tau )$, which is,
\begin{equation}\label{toftau}
t(\tau )=\frac{2 \left(3 H_0 t_i+\,\,\ln\left(\frac{\left(6 H_0+M^2 t_i\right) (\tau -\tau_0)}{2 a_0^3}\right)\right)}{6 H_0+M^2 t_i}\, ,
\end{equation}
with $\tau_0$ being a fiducial initial time instance. Having Eq.~(\ref{toftau}), we can express the scale factor and the Hubble rate as functions of $\tau$, so the scale factor $a(\tau)$ is equal to,
\begin{equation}\label{scaletaufactor}
a(\tau)= \frac{\left(\left(6 H_0+M^2 t_i\right) (\tau -\tau_0)\right)^{1/3}}{2^{1/3}}\, ,
\end{equation}
while the Hubble rate $K(\tau)$ becomes equal to,
\begin{equation}\label{khubble}
K(\tau )=\frac{1}{3 (\tau -\tau_0)}\, .
\end{equation}
In addition, the scalar curvature as a function of the coordinate $\tau$, that is, $R(\tau)$, is equal to,
\begin{equation}\label{scalarcurvatau}
R=\frac{\left(6 H_0+M^2 t_i\right)^{8/3} (\tau -\tau_0)^{2/3}}{3\cdot 2^{2/3}}\, ,
\end{equation}
which can be inverted to yield the function $\tau(R)$, which is,
\begin{equation}\label{taurho}
\tau-\tau_0=\frac{2\cdot 3^{3/2}}{ \left( 6H_0+M^2t_i \right)^4}R^{3/2}\, .
\end{equation}
The above relation (\ref{taurho}) will be useful for the determination of the $F(R)$ gravity, as we now demonstrate. Let us see which vacuum $F(R)$ gravity generates the evolution (\ref{scaletaufactor}), and by combining Eqs.~(\ref{khubble}) and (\ref{Uni16}), with $\rho=p=0$, we obtain the following differential equation,
\begin{equation}\label{diffeqnoftau}
\frac{ d^2\sigma}{ d\tau^2}+\frac{2}{3(\tau-\tau_0)}\frac{ d\sigma}{ d\tau}=0\, ,
\end{equation}
where we denoted $\sigma=F'(R)$, and the prime denotes differentiation with respect to the Ricci scalar $R$. The differential equation (\ref{diffeqnoftau}) can easily be solved to yield the function $\sigma(\tau)$ which is,
\begin{equation}\label{simgtau}
\sigma (\tau)=3 (\tau-\tau_0)^{1/3} C_1+C_2\, ,
\end{equation}
with $C_1$ and $C_2$ being arbitrary integration constants. By using Eq.~(\ref{taurho}), the function $\sigma(R)=F'(R)$ becomes,
\begin{equation}\label{frdot}
F'(R)=\frac{3^{3/2}2^{1/3}}{(6H_0+M^2t_i)^{4/3}}R^{1/2}C_1+C_2\, ,
\end{equation}
so by integrating with respect to $R$, we obtain the final form of the $F(R)$ gravity which generates the unimodular gravity model (\ref{scaletaufactor}), which is,
\begin{equation}\label{finalfrfunction}
F(R)=R+\frac{3^{1/2}2^{4/3}}{(6H_0+M^2t_i)^{4/3}}R^{3/2}\, ,
\end{equation}
where we have set the arbitrary integration constant $C_2=1$. Moreover, we can explicitly calculate the function $\lambda$ by using 
Eq.~(\ref{Uni14}), and also (\ref{frdot}), with $\rho=0$ and $C_2=1$, and the resulting expression for the unimodular Lagrange multiplier $\lambda$ is,
\begin{align}\label{finallambdaoftau}
& \lambda (\tau)= - \frac{\left(6+\mathcal{Q}^2 (-1+3 \tau -3 \tau_0)\right) \left(\mathcal{Q}^{4/3}+3\ 2^{1/9} 3^{1/6} C_1 \left(a_0^2 \mathcal{Q}^2 \left(\frac{\mathcal{Q} (\tau -\tau_0)}{a_0^3}\right)^{2/3}\right)^{1/3}\right)}{6 \mathcal{Q}^{4/3}}\, .
\end{align}
where $\mathcal{Q}=(6H_0+M^2t_i)$. 

We can have a rough estimate for the observational implications of the model (\ref{scaletaufactor}), for early times. To this end we shall compute the Hubble slow-roll indices 
\cite{Liddle:1994dx,Liddle:1992wi,Copeland:1993jj} for the model (\ref{scaletaufactor}), but we need to note that in order to extract a clear picture regarding the primordial curvature perturbations, the explicit calculation of the comoving curvature perturbation for the $F(R)$ model at hand is needed. However, we can have a rough estimate for the observational indices for the case at hand too. The Hubble slow-roll indices are defined in terms of the cosmic time $t$, as follows,
\begin{equation}\label{hubbleslowrooll}
\epsilon_H=-\frac{\dot{H}}{H^2}\, ,\quad \eta_H=-\frac{\ddot{H}}{2H\dot{H}}\, ,
\end{equation}
so by using the variable $\tau$, we can express the Hubble slow-roll parameters as functions of the variable $\tau$ as follows,
\begin{equation}\label{hubbleslowrooll2}
\epsilon_H=-\frac{\frac{ dH}{ d\tau}}{H(\tau)^2\frac{ dt}{ d\tau}} \, , \quad \eta_H=-\frac{\frac{ d^2H}{ d\tau^2}\left(\frac{ dt}{ d\tau}\right)^2-\left(\frac{ dH}{ d\tau}\right)\frac{ d^2t}{ d\tau}^2(\frac{ dt}{ d\tau})^{-1}}{2H(\tau)\frac{ dH}{ d\tau}\frac{ dt}{ d\tau}}\, ,
\end{equation}
and by using the relation  (\ref{toftau}), the Hubble slow-roll parameter $\epsilon_H$ becomes,
\begin{equation}\label{hubbleslowrollparameters1}
\epsilon_H=\frac{12 \left(6 H_0 M+M^3 t_i\right)^2}{\left(36 H_0^2+M^4 t_i^2-4 M^2 \,\,\ln\left(\frac{\left(6 H_0+M^2 t_i\right) (\tau -\tau_0)}{2 a_0^3}\right)\right)^2}\, ,
\end{equation}
while $\eta_H$ takes the following form,
\begin{equation}\label{hubbleslowrollparameters12}
\eta_H=-\frac{3 \left(-4+36 H_0^2 (\tau -\tau_0)^2+12 H_0 M^2 t_i (\tau -\tau_0)^2+M^4 t_i^2 (\tau -\tau_0)^2\right)}{2 (\tau -\tau_0)^2 \left(36 H_0^2+M^4 t_i^2-4 M^2 \,\ln\left(\frac{\left(6 H_0+M^2 t_i\right) (\tau -\tau_0)}{2 a_0^3}\right)\right)}\, .
\end{equation}
Having the slow-roll indices (\ref{hubbleslowrollparameters1}), (\ref{hubbleslowrollparameters12}) at hand, we can compute the spectral index of primordial curvature perturbations, which in terms of the Hubble slow-roll parameters is given as follows,
\begin{equation}\label{spectralindexhubslow}
n_s=1-4\epsilon_H+2\eta_H\, ,
\end{equation}
but in order to obtain an exact value of the spectral index of primordial curvature perturbations, it is more useful to use the $e$-foldings number, instead of the variable $\tau$, so we use the definition of the $e$-foldings number, which is,
\begin{equation}\label{efolding}
N=\int_{t_i}^tH(t) dt\, .
\end{equation}
So by using (\ref{efolding}) and (\ref{toftau}), we can find an explicit relation between $\tau$ and $N$, which is,
\begin{equation}\label{ntauefolidnfg}
\tau-\tau_0=\frac{2 \e^{3 \left(6 H_0^2 t_i+M^2 N t_i+H_0 \left(6 N+t_i \left(-1+M^2 t_i\right)\right)\right)}}{6 H_0+M^2 t_i}\, ,
\end{equation}
and by substituting in the Hubble slow-roll parameters, the spectral index of primordial curvature perturbations takes the following form,
\begin{align}\label{spprmcb}
n_s=&1-\frac{48 \left(6 H_0 M+M^3 t_i\right)^2}{\left(36 H_0^2+M^4 t_i^2-4 M^2 \,\ln\left(\frac{ \e^{3 \left(6 H_0^2 t_i+M^2 N t_i+H_0 \left(6 N+t_i \left(-1+M^2 t_i\right)\right)\right)}}{a_0^3}\right)\right)^2} \nn 
& +\frac{3 \e^{-6 (N+H_0 t_i) \left(6 H_0+M^2 t_i\right)} \left( \e^{6 H_0 t_i}-\e^{6 (N+H_0 t_i) \left(6 H_0+M^2 t_i\right)}\right) \left(6 H_0+M^2 t_i\right)^2}{36 H_0^2+M^4 t_i^2-4 M^2 \,\ln\left(\frac{ \e^{3 \left(6 H_0^2 t_i+M^2 N t_i+H_0 \left(6 N+t_i \left(-1+M^2 t_i\right)\right)\right)}}{a_0^3}\right)}\, .
\end{align}
By using the following values for the set of the free parameters $H_0$, $N$, $M$, $a_0$, and $t_i$,
\begin{equation}\label{valuespar}
H_0=0.00000016\, , \quad M=0.053\, , \quad t_i=10^{-43}\mathrm{sec}\, , \quad a_0=1\, ,
\end{equation}
we obtain that $n_s\simeq 0.966$, which is compatible with the Planck result for the spectral index, which is,
\be
\label{planckconstr}
n_s=0.9644\pm 0.0049\, .
\ee
We need to stress however that the observational quantities have to be evaluated by calculating the comoving curvature perturbation and the corresponding evolution of the curvature perturbations, in order to provide a more solid result. We intend to present the result of this calculation in a future work.

\subsection{Unimodular $F(R)$ Gravity as Perfect Fluid and Slow-roll Inflation}

A quite convenient way of studying general $F(R)$ theories of gravity, which enables us to reveal the slow-roll inflation evolution of a specific cosmological evolution, is by treating the $F(R)$ gravity cosmological system as a perfect fluid. This approach was developed 
in Ref.~\cite{Bamba:2014wda}, and as was evinced, the slow-roll indices and the corresponding observational indices receive quite convenient form, and the study of the inflationary evolution is simplified to a great extent. Actually, the slow-roll indices and the corresponding inflationary indices can be expressed in terms of the Hubble rate, see Ref.~\cite{Bamba:2014wda} (and also \cite{Mukhanov:2014uwa}). Particularly, for a given cosmological evolution with Hubble rate $H(N)$, the slow-roll indices $\epsilon$, $\eta$ can be written in terms of the Hubble rate $H(N)$ as follows,
\begin{align}
\label{S7}
\epsilon
=& - \frac{ H(N)}{4  H'(N)} \left( \frac{6\frac{ 
H'(N)}{ H(N)}
+ \frac{ H''(N)}{ H(N)} + \left( \frac{ H'(N)}{ 
H(N)} \right)^2}
{3 + \frac{ H'(N)}{ H(N)}} \right)^2 \, , \nn
\eta = & -\frac{1}{2} \left( 3 + \frac{ H'(N)}{ H(N)} \right)^{-1} 
\left(
9 \frac{ H'(N)}{ H(N)} + 3 \frac{ H''(N)}{ H(N)}
+ \frac{1}{2} \left( \frac{ H'(N)}{ H(N)} \right)^2 -\frac{1}{2} 
\left( \frac{ H''(N)}{ H'(N)} \right)^2
+ 3 \frac{ H''(N)}{ H'(N)} + \frac{ H'''(N)}{ H'(N)} 
\right) \, , \nn
\xi^2 = & \frac{ 6 \frac{ H'(N)}{ H(N)} + \frac{ 
H''(N)}{ H(N)}
+ \left( \frac{ H'(N)}{ H(N)} \right)^2 }{4 \left( 3 + \frac{ 
H'(N)}{ H(N)} \right)^2}
\left( 3 \frac{ H(N)  H'''(N)}{ H'(N)^2} + 9 \frac{ 
H'(N)}{ H(N)}
 - 2 \frac{ H(N)  H''(N)  H'''(N)}{ H'(N)^3} + 4 
\frac{ H''(N)}{ H(N)}
\right. \nn
& \left.
+ \frac{ H(N)  H''(N)^3}{ H'(N)^4} + 5 \frac{ 
H'''(N)}{ H'(N)}
 - 3 \frac{ H(N)  H''(N)^2}{ H'(N)^3} - \left( \frac{ 
H''(N)}{ H'(N)} \right)^2
+ 15 \frac{ H''(N)}{ H'(N)}
+ \frac{ H(N)  H''''(N)}{ H'(N)^2} \right)\, ,
\end{align}
where $N$ is the $e$-folding number, which is related to the scale factor $a(t)$ as, $\frac{a}{a_0}=\e^N$. Consider the case in which, $f_0=\frac{1}{3}$, which corresponds to the de Sitter spacetime, because we are now interested in the slow-roll inflation regime. Then, by using Eqs.~(\ref{Uni2}) and (\ref{Uni25}), we find,
\be
\label{Uni32}
H \equiv \frac{1}{a}\frac{da}{dt} = \frac{1}{a} \frac{d\tau}{dt} \frac{da}{d\tau} = a^2 \frac{da}{d\tau}
=H_0 + 3 H_0 \left( b(\tau) + \tau \frac{d b(\tau)}{d\tau} \right)\, ,
\ee
where the parameter $H_0$ satisfies,
\be
\label{Uni33}
H_0 \equiv \frac{1}{3\tau_0}\, .
\ee
Consequently, owing to the fact that $\frac{dN}{d\tau} = K$, we find,
\begin{align}
\label{Uni34}
H'(N) =& 9H_0 \left( 2 \tau \frac{d b(\tau)}{d\tau} + \tau^2 \frac{d^2 b(\tau)}{d\tau^2} \right) \, , \quad 
H''(N) = 27H_0 \left( 2 \tau \frac{d b(\tau)}{d\tau} + 4 \tau^2 \frac{d^2 b(\tau)}{d\tau^2}
+ \tau^3 \frac{d^3 b(\tau)}{d\tau^3} \right) \, , \nn 
H'''(N) =& 81H_0 \left( 2 \tau \frac{d b(\tau)}{d\tau} + 10 \tau^2 \frac{d^2 b(\tau)}{d\tau^2}
+ 7 \tau^3 \frac{d^3 b(\tau)}{d\tau^3} + \tau^4 \frac{d^4 b(\tau)}{d\tau^4} \right) \, , \nn 
H''''(N) =& 243H_0 \left( 2 \tau \frac{d b(\tau)}{d\tau} + 22 \tau^2 \frac{d^2 b(\tau)}{d\tau^2}
+ 31 \tau^3 \frac{d^3 b(\tau)}{d\tau^3} + 11 \tau^4 \frac{d^4 b(\tau)}{d\tau^4} 
+  \tau^5 \frac{d^5 b(\tau)}{d\tau^5} \right) \, ,
\end{align}
and therefore, the corresponding slow-roll indices read,  
\begin{align}
\epsilon =& \frac{81 \tau \left( 4 \frac{d b(\tau)}{d\tau} + 4 \tau \frac{d^2 b(\tau)}{d\tau^2}
+ \tau^2 \frac{d^3 b(\tau)}{d\tau^3} \right)^2}{4  \left( 2 \frac{d b(\tau)}{d\tau} + \tau \frac{d^2 b(\tau)}{d\tau^2} \right)} \, , \nn
\eta = & \frac{3}{4} \left( \frac{2 \frac{d b(\tau)}{d\tau} + 4 \tau \frac{d^2 b(\tau)}{d\tau^2}
+ \tau^2 \frac{d^3 b(\tau)}{d\tau^3}}{ 2 \frac{d b(\tau)}{d\tau} + \tau \frac{d^2 b(\tau)}{d\tau^2} }\right)^2 
 - \frac{3 \left( 4 \frac{d b(\tau)}{d\tau} + 14 \tau \frac{d^2 b(\tau)}{d\tau^2}
+ 8 \tau^2 \frac{d^3 b(\tau)}{d\tau^3} + \tau^3 \frac{d^4 b(\tau)}{d\tau^4} \right)}{2 \left(
 2 \frac{d b(\tau)}{d\tau} + \tau \frac{d^2 b(\tau)}{d\tau^2} \right)} \, .
\end{align}

In the perfect fluid approach of Ref.~\cite{Bamba:2014wda}, the spectral index of primordial curvature perturbations $n_s$ and the scalar-to-tensor ratio $r$ can be expressed in terms of the slow-roll parameters (\ref{S7}) as follows,
\begin{equation}\label{indexspectrscratio}
n_s\simeq 1-6 \epsilon +2\eta\, , \quad r=16\epsilon \, .
\end{equation}
We need to stress that the approximations for the observational indices $n_s$ and $r$, remain valid if for a wide range of values of the $e$-foldings number $N$, the slow-roll indices satisfy $\epsilon,\eta \ll 1$. We shall now present a model with quite interesting phenomenology, and we shall investigate which unimodular $F(R)$ gravity can generate this model. Before starting, recall that the recent Planck data \cite{Ade:2015lrj,Planck:2013jfk} indicate that the spectral index $n_s$ and the scalar-to-tensor ratio, are constrained as follows,
\begin{equation}\label{constraintedvalues}
n_s=0.9644\pm 0.0049\, , \quad r<0.10\, ,
\end{equation}
while the most recent BICEP2 data \cite{Array:2015xqh} further constrain $r$ to be $r<0.07$. Having these in mind, consider the cosmological evolution with the following Hubble rate as a function of the $e$-folding number,
\begin{equation}\label{hub2}
H(N)=\left(-\gamma\, \e^{\delta N }+\zeta\right)^b\, .
\end{equation}
Substituting the Hubble rate (\ref{hub2}) in the slow-roll parameters (\ref{S7}), these become,
\begin{align}\label{hubslowroll2}
\epsilon=& -\frac{b \e^{\delta N} \gamma \delta  \left(\zeta (6+\delta )-2 \e^{\delta N} \gamma (3+b \delta )\right)^2}{4 \mathcal{G}(N)} \\
\label{edggs}
\eta = &-\frac{\delta  \left(8 b^2 \e^{2\delta N} \gamma^2 \delta +\zeta \left(2 \e^{\delta N} \gamma (-3+\delta )+\zeta (6+\delta )\right)+2 b \e^{\delta N} \gamma \left(12 \e^{\delta N} \gamma-\zeta (12+5 \delta )\right)\right)}{4 \left( \e^{\delta N} \gamma-\zeta\right) \left(-3 \zeta+ \e^{\delta N} \gamma (3+b \delta )\right)}\, ,
\end{align}
where we introduced the function $\mathcal{G}(N)$, which is equal to,
\begin{equation}\label{hfghfhgfghdf}
\mathcal{G}(N)=\left( \e^{\delta N} \gamma-\zeta\right) \left(-3 \zeta+ \e^{\delta N} \gamma (3+b \delta )\right)^2 \, .
\end{equation}
Having at hand Eqs.~(\ref{hubslowroll2}) and (\ref{edggs}), the calculation of the observational indices can easily be done, and the spectral index $n_s$ reads,
\begin{align}
\label{scalarpertandsctotenso}
n_s=& \frac{2 \left( \e^N\right)^{3 \delta } \gamma^3 (3+b \delta )^2 (1+2 b \delta )+3 \zeta^3 \left(-6+6 \delta +\delta ^2\right)}{2 \mathcal{G}(N)}
+\frac{ \e^{\delta N} \gamma \zeta^2 \left(54+12 (-3+4 b) \delta +3 \delta ^2+2 b \delta ^3\right)}{2 \mathcal{G}(N)} \nn
& -\frac{2 \e^{2\delta N} \gamma^2 \zeta \left(27+(-9+48 b) \delta +\left(3+13 b^2\right) \delta ^2+b (1+b) \delta ^3\right)}{2 \mathcal{G}(N)}
\, ,
\end{align}
while the scalar-to-tensor ratio $r$ has the following form,
\begin{equation}\label{thodorakis}
r=-\frac{4 b \e^{\delta N} \gamma \delta  \left(\zeta (6+\delta )-2 \e^{\delta N} \gamma (3+b \delta )\right)^2}{\mathcal{G}(N)} \, .
\end{equation}
Concordance with observations can be achieved if we appropriately choose the parameters $\gamma$, $\zeta$, $\delta$, and $b$, so by making the following choice,
\begin{equation}\label{parmchoice12}
\gamma=0.5\, , \quad \zeta=10\, , \quad \delta=\frac{1}{48}\, , \quad b=1\, ,
\end{equation}
the observational indices $n_s$ and $r$, take the following values,
\begin{equation}\label{indnewparadigm12}
n_s\simeq 0.965735\, , \quad r=0.0554765\, ,
\end{equation}
which are compatible with both the latest Planck data \cite{Ade:2015lrj,Planck:2013jfk} and the latest BICEP2 data \cite{Array:2015xqh}.

Let us now investigate which unimodular $F(R)$ gravity can generate the cosmological evolution (\ref{hub2}), and in order to do so, we need to investigate which scale factor corresponds to the Hubble rate (\ref{hub2}). Since, $H=\dot{a}/a$, $\e^N=a/a_0$, $b=1$ and also by using 
Eq.~(\ref{Uni2}) and assuming $a_0=1$, without loss of generality, we have,
\begin{equation}\label{neweqns1}
\frac{ da}{a(-\gamma a^{\delta}+\zeta)}=\frac{ d\tau}{a^3}\, ,
\end{equation}
which can be written as follows,
\begin{equation}\label{neweqns2}
\frac{a^2 da}{(-\gamma)a^{\delta}+\zeta}= d\tau\, .
\end{equation}
By substituting the variable $\delta=1/48$ which we used in Eq.~(\ref{parmchoice12}), and by integrating Eq.~(\ref{neweqns2}) by parts, we obtain the following quite lengthy result, where we kept the first and last terms,
\begin{align}\label{extensive}
&\tau-\tau_0=-\frac{48 a^{1/48} \zeta^{142}}{\gamma^{143}}\ldots-\frac{48 a^{143/48}}{143 \gamma}-\frac{24 a^{71/24} \zeta}{71 \gamma^2}-\frac{16 a^{47/16} \zeta^2}{47 \gamma^3}-\frac{48 \zeta^{143}
\,\ln\left(a^{1/48} \gamma-\zeta\right)}{\gamma^{144}}\, ,
\end{align}
where $\tau_0$ some fiducial initial time. Solving Eq.~(\ref{extensive}), with respect to $a$ 
is a rather formidable task, so since we are interested in early-time evolution, the scale factor satisfies $a<1$, so we keep only the first term in the above expression, and eventually we have,
\begin{equation}\label{fracofgy}
\tau-\tau_0= -\frac{48 a^{143/48}}{143 \gamma}\, ,
\end{equation}
which can be solved with respect to $a$, and hence the scale factor as a function of $\tau$ reads,
\begin{equation}\label{approatau}
a(\tau)\simeq \mathcal{A}_1 (\tau-\tau_0)^{48}\, ,
\end{equation}
where the parameter $\mathcal{A}_1$ is,
\begin{equation}\label{auxiliary}
\mathcal{A}_1=\frac{\zeta^{143/48}}{48^{48}\gamma^{142/48}}\, .
\end{equation}
The Hubble rate $K(\tau)$ and the Ricci scalar $R$ for the scale factor (\ref{approatau}) read,
\begin{equation}\label{metapomenon}
K(\tau)=\frac{48}{\tau -\tau_0}\, , \quad R(\tau)=68832 \mathcal{A}_1^8 (\tau -\tau_0)^{382}\, ,
\end{equation}
and the differential equation (\ref{Uni16}) becomes,
\begin{equation}\label{diffeneforlastparadigm}
\frac{13728 }{(\tau -\tau_0)^2}y(\tau )+\frac{96 }{\tau -\tau_0}\dot{y}(\tau )+\ddot{y}(\tau )=0\, ,
\end{equation}
where $y=F'(R(\tau))$ and the ``dot'' denotes differentiation with respect to $\tau$. Solving the differential equation (\ref{diffeneforlastparadigm}), we obtain the following solution,
\begin{equation}\label{dksaenypsististheo}
y(\tau )=\frac{c_2 \cos \left(\frac{1}{2} \sqrt{45887} \,\ln[\tau -\tau_0]\right)}{(\tau -\tau_0)^{95/2}}+\frac{c_1 \sin \left(\frac{1}{2} \sqrt{45887} \,\,\ln[\tau -\tau_0]\right)}{(\tau -\tau_0)^{95/2}}\, ,
\end{equation}
with $c_1$ and $c_2$ being arbitrary integration constants. From Eq.~(\ref{metapomenon}), we can obtain the function $\tau=\tau (R)$, so by substituting the resulting expression in Eq. (\ref{dksaenypsististheo}), we obtain the function $F'(R)$, which is,
\begin{equation}\label{lastequat}
F'(R)=\frac{2^{475/764} 3^{95/382} 239^{95/764} \left(c_2 \cos \left(\frac{1}{764} \sqrt{45887} \,\,\ln\left(\frac{R}{68832 \mathcal{A}_1^8}\right)\right)+c_1 \sin \left(\frac{1}{764} \sqrt{45887} \,\,\ln\left(\frac{R}{68832 \mathcal{A}_1^8}\right)\right)\right)}{\left(\frac{R}{\mathcal{A}_1^8}\right)^{95/764}}\, .
\end{equation} 
Note that in such models of unimodular $F(R)$ gravity, graceful exit from inflation may be achieved either via the contribution of $R^2$ correction terms, or via a Type IV singularity, in which case singular inflation might occur.

\section{The Newton Law in Unimodular $F(R)$ Gravity}

In this section we shall discuss an important implication of the Unimodular $F(R)$ gravity formalism and particularly the Newton law in the case of unimodular $F(R)$ gravity.

The total unimodular $F(R)$ gravity with matter fluids action is written as follows, 
\be
\label{UF2}
S = \int d^4 x \left\{ \sqrt{-g} \left( \frac{F(R)}{2\kappa^2} - \lambda \right) + \lambda \right\} 
+ S_\mathrm{matter} \left( g_{\mu\nu}, \Psi \right)\, ,
\ee
with the action $S_\mathrm{matter}$ denoting all the matter fluids present, and $\Psi$ denotes the matter fluids. The scalar-tensor counterpart theory is,
\be
\label{UF3}
S = \int d^4 x \left\{ \sqrt{-g} \left( \frac{1}{2\kappa^2} \left( R 
- \frac{3}{2} g^{\mu\nu} \partial_\mu \phi \partial_\nu \phi - V(\phi) \right)- \lambda \e^{2\phi} \right) + \lambda \right\} 
+ S_\mathrm{matter} \left( \e^\phi g_{\mu\nu}, \Psi \right)\, ,
\ee
with the potential $V(\phi)$ being equal to,  
\be
\label{UF4}
V(\phi) = \frac{A(\phi)}{F'\left( A \left( \phi \right) \right)} 
 - \frac{F\left( A \left( \phi \right) \right)}{F'\left( A \left( \phi \right) \right)^2}\, ,
\ee
and also the function $A(\phi)$ can be found by solving, 
\be
\label{UF5}
\phi = - \ln F'(A)\, .
\ee
In the Einstein frame, the unimodular constraint (\ref{Uni1}) becomes,
\be
\label{UF6}
\e^{2\phi}\sqrt{-g} = 1 \, .
\ee
Hence by eliminating the scalar field $\phi$ we get, 
\be
\label{UF7}
S = \int d^4 x \sqrt{-g} \left( \frac{1}{2\kappa^2} \left( R - \frac{3}{32 g^2} 
g^{\mu\nu} \partial_\mu g \partial_\nu g - V\left(\frac{1}{4}\ln \left( - g \right) \right) \right) 
\right) 
+ S_\mathrm{matter} \left( \left( -g \right)^\frac{1}{4} g_{\mu\nu}, \Psi \right)\, .
\ee
We now perform a Newtonian approximation of the theory, so we perturb the metric as follows, 
\be
\label{UF8}
g_{\mu\nu} = g_{\mu\nu}^{(0)} + h_{\mu\nu}  \, .
\ee
By assuming that the background metric is flat, that is, $g_{\mu\nu}^{(0)} = \eta_{\mu\nu}$, the scalar curvature reads,
\be
\label{UF10}
\sqrt{- g} R \sim - \frac{1}{2} \partial_\lambda h_{\mu\nu} \partial^\lambda h^{\mu\nu} 
+ \partial_\lambda h^\lambda_{\ \mu} \partial_\nu h^{\mu\nu} 
 - \partial_\mu h^{\mu\nu} \partial_\nu h + \frac{1}{2}\partial_\lambda h \partial^\lambda h  \, ,
\ee
with $h$ denoting the trace of the perturbation metric $h_{\mu\nu}$, $h \equiv \eta^{\rho\sigma} h_{\rho\sigma}$. Since we assumed a flat background, we get,
\be
\label{UF10b}
V(0)=V'(0)=0\, ,
\ee
and the potential $V$ can be approximated as follows,  
\be
\label{UF10c}
V \sim \frac{1}{2}m^2 h^2\, .
\ee
Finally, the linearized action reads,
\begin{align}
\label{UF12}
S =& \frac{1}{2\kappa^2} \int d^4 x \left\{ 
  - \frac{1}{2} \partial_\lambda h_{\mu\nu} \partial^\lambda h^{\mu\nu} 
+ \partial_\lambda h^\lambda_{\ \mu} \partial_\nu h^{\mu\nu} 
 - \partial_\mu h^{\mu\nu} \partial_\nu h + \frac{1}{2}\partial_\lambda h \partial^\lambda h  
 - \frac{3}{32}\partial_\mu h \partial^\mu h - \frac{1}{2}m^2 h^2 \right\} \nn
& + S_\mathrm{matter} \left( \eta_{\mu\nu} + h_{\mu\nu} - \frac{1}{4}\eta_{\mu\nu} h , \Psi \right)\, .
\end{align}
Upon variation of the action (\ref{UF12}) with respect to the metric $h_{\mu\nu}$, we get,
\be
\label{UF13}
\partial_\lambda \partial^\lambda h_{\mu\nu} 
 - \partial_\mu \partial^\lambda h_{\lambda\nu}
 - \partial_\nu \partial^\lambda h_{\lambda\mu} 
+ \partial_\mu \partial_\nu h 
+ \eta_{\mu\nu} \partial^\rho \partial^\sigma h_{\rho\sigma} 
 - \frac{13}{16} \eta_{\mu\nu} \partial_\lambda \partial^\lambda h  - m^2 \eta_{\mu\nu} h 
= \kappa^2 \left( T_{\mu\nu} - \frac{1}{4} \eta_{\mu\nu} T \right) \, ,
\ee
where $T_{\mu\nu}$ is the energy-momentum tensor and $T$ is the trace of 
$T_{\mu\nu}$, with $T \equiv \eta^{\rho\sigma} T_{\rho\sigma}$. Also by multiplying Eq.~(\ref{UF13}) with $\eta^{\mu\nu}$, we get,
\be
\label{UF13b}
0 = - \frac{5}{4} \partial_\lambda \partial^\lambda h - 4 m^2 h + 2 \partial^\mu \partial^\nu h_{\mu\nu}\, .
\ee
Consider the point-like source at the origin, with the following components of the energy-momentum tensor,
\be
\label{UF14}
T_{00} = M \delta \left( r \right)\, , \quad T_{ij} = 0 \, \left(i,j=1,2,3\right) \, .
\ee
By considering only static solutions, the $(0,0)$, $(i,j)$, and $(0,i)$ components of Eq.~(\ref{UF13}) and Eq.~(\ref{UF13b}) are,
\begin{align}
\label{UF15}
& \partial_i \partial^i h_{00} - \partial^i \partial^j h_{ij} + \frac{13}{16} \partial_i \partial^i h  
+ m^2 h = \frac{3 \kappa^2}{4} M \delta \left( r \right) \, , \\
\label{UF16}
& \partial_k \partial^k h_{ij} - \partial_i \partial^k h_{kj} - \partial_j \partial^k h_{ki} 
+ \partial_i \partial_j h + \delta_{ij} \partial^k \partial^l h_{kl} 
 - \frac{13}{16} \delta_{ij} \partial_k \partial^k h  - m^2 \delta_{ij} h 
= \frac{\kappa^2}{4} M \delta \left( r \right) \, , \\
\label{UF17}
& \partial_j \partial^j h_{0i} - \partial_i \partial^k h_{k0} = 0 \, ,\\
\label{UF18}
& - \frac{5}{4} \partial_k \partial^k h - 4 m^2 h + 2 \partial^i \partial^j h_{ij} = 0 \, .
\end{align}
Notice that in contrast to the usual Einstein-Hilbert gravity, where four gauge degrees of the freedom exist, in the case of unimodular $F(R)$ gravity, only three gauge degrees of freedom exist, due to the unimodular constraint (\ref{Uni1}). By imposing the following gauge conditions, 
\be
\label{UF19}
\partial^i h_{ij} = 0 \, .
\ee
Eq.~(\ref{UF18}) now becomes, 
\be
\label{UF20}
 - \frac{5}{4} \partial_k \partial^k h - 4 m^2 h = 0 \, ,
\ee
and by choosing the boundary condition appropriately we get,
\be
\label{UF21}
h=0 \, .
\ee
Hence, by combining Eqs. (\ref{UF19}) and (\ref{UF21}), Eqs.~(\ref{UF15}), (\ref{UF16}), 
and (\ref{UF17}) can be rewritten as follows, 
\begin{align}
\label{UF22}
& \partial_i \partial^i h_{00} = \frac{3 \kappa^2}{4} M \delta \left( r \right) \, , \\
\label{UF23}
& \partial_k \partial^k h_{ij} = \frac{\kappa^2}{4} M \delta \left( r \right) \, , \\
\label{UF24}
& \partial_j \partial^j h_{0i} - \partial_i \partial^k h_{k0} = 0 \, .
\end{align}
Also by appropriately choosing the boundary conditions, the equations above, in conjunction with Eq. (\ref{UF21}) become, 
\be
\label{UF25}
h_{0i}=0\, , \quad h_{ij} = \frac{1}{3} \delta_{ij} h_{00}\, .
\ee
The Newtonian gravitational potential $\Phi$ is defined by $h_{00} = 2 \Phi$, and therefore Eq.~(\ref{UF22}) yields the Poisson equation for 
the Newtonian potential $\Phi$, which is,
\be
\label{UF26}
\partial_i \partial^i \Phi = \frac{3 \kappa^2}{8} M \delta \left( r \right) \, .
\ee
In effect, by redefining the gravitational constant $\kappa$ as follows, 
\be
\label{UF27}
\frac{3 \kappa^2}{4} \to \kappa^2 = 8\pi G \, ,
\ee
we get the standard Poisson equation satisfied by the Newtonian potential $U$,
\be
\label{UF28}
\partial_i \partial^i \Phi = 4\pi G M \delta \left( r \right) \, ,
\ee
with the solution being equal to, 
\be
\label{UF29}
\Phi = - \frac{GM}{r}\, .
\ee
In conclusion, the resulting Newtonian limit of the unimodular $F(R)$ gravity is different from the usual $F(R)$ gravity case, where corrections to Newton's law appear.

\section{Conclusions}

In this paper we studied the extension of general relativity unimodular gravity, in the context of $F(R)$ gravity. In the standard unimodular gravity, the determinant of the metric is constrained to be a fixed number or a function, so we investigated which metric can be compatible with such a constraint, since the FRW metric cannot be compatible. After discussing which metric can be compatible with the unimodular constraint, we imposed this constraint in a Jordan frame $F(R)$ gravity, and we derived the equations of motion, by employing the metric formalism. The resulting equations constitute a reconstruction method which can determine which unimodular $F(R)$ gravity can realize some given cosmological evolution. Also, it enables us to find which cosmological evolution corresponds to a given $F(R)$ gravity. The resulting picture is different from the ordinary $F(R)$ gravity and also different from unimodular Einstein-Hilbert gravity, as was probably expected. By using some characteristic examples, we demonstrated how the reconstruction method operates, and we investigated some inflation-related paradigms, for which we calculated the corresponding Hubble slow-roll parameters. Also, we gave a rough estimate of the observational indices but the calculation has to be done by using the comoving curvature perturbation, and investigating if the resulting evolution of this gauge-invariant quantity leads to a scale-invariant spectrum. This is because, although the unimodular gravity and the Einstein-Hilbert gravity are identical at large scales, this is not necessarily so for the case of unimodular gravity, so this has to be explicitly demonstrated by investigating the perturbation equations from scratch. A way to circumvent this calculation is to treat the unimodular $F(R)$ gravity system as a perfect fluid, in which case the slow-roll parameters and the corresponding observational indices, can easily be calculated. We performed this calculation and we demonstrated that in some specific cases, concordance with the observational data can be achieved.  

The potential applications of the reconstruction method we presented in this paper are quite many, since it is possible to realize various cosmological evolutions, which are exotic for the standard Einstein-Hilbert gravity and cannot be realized in that case, for example bouncing cosmologies \cite{Novello:2008ra,Li:2014era,Brandenberger:1993ef,Mukhanov:1991zn,Cai:2013kja,Cai:2014zga,deHaro:2014kxa,Qiu:2010ch,Qiu:2010vk,
Lehners:2011kr,Khoury:2012dn,Koehn:2013upa,Easson:2011zy,Cai:2012va,Odintsov:2015zua,Lehners:2015mra,Cai:2007qw}. Also, the unimodular $F(R)$ gravity formalism can be extended in other modified gravity theories, with Lagrangian $L= \sqrt(-g)\left(F(R, R^{\mu \nu}R_{\mu \nu} , R_{ \mu \nu \alpha \beta} R^{ \mu \nu \alpha \beta} \right)-\lambda)+\lambda$, or for non-local gravity, with Lagrangian $L=\sqrt(-g)\left(F(R,R\square^m R,\square^d R)-\lambda\right)+\lambda $, e.t.c., where the coefficients $m$ and $d$ can be positive and/or negative. Another potentially interesting application of unimodular F(R) gravity, and of ordinary Einstein-Hilbert unimodular gravity, is to investigate the solutions corresponding to relativistic stars
and black holes. Some important deviations from the general relativistic predictions for relativistic stars are
obtained for certain $F(R)$ gravities, so it would be interesting to investigate what new issues the unimodular gravity
and unimodular $F(R)$ gravity bring along. Also, since at large scales, the ordinary Einstein-Hilbert gravity and it's
unimodular counterpart are indistinguishable, are these two theories indistinguishable at the level of
compact objects? And what is the role of unimodular $F(R)$ gravity in the cases that compact objects are considered.
Also, an issue we did not address in this paper is the Einstein frame counterpart of the unimodular $F(R)$ gravity
theory, which has to be thoroughly investigated, and especially to see how the corresponding scalar-tensor theory
behaves near Big Rip singularities and other finite-time singularities \cite{Nojiri:2005sx}. Particularly, the dynamical behavior of
cosmological systems near mild singularities, like the Type IV singularity, is strongly affected, as was shown in
\cite{Barrow:2015ora,Nojiri:2015fra,Oikonomou:2015qha,Kohli:2015fva,Oikonomou:2015qfh,Nojiri:2015qyc,Odintsov:2015jca,Odintsov:2015zza,Odintsov:2015gba}. Therefore, a concrete analysis of the unimodular $F(R)$ gravity theory near mild and even crushing types
singularities is of some interest, and should be appropriately addressed in a future work. We hope to address some of
the aforementioned issues in a future work.

\section*{Acknowledgments}

This work is supported by MINECO (Spain), project
  FIS2013-44881 and I-LINK 1019 (S.D.O) and by Min. of Education and 
Science of Russia 
(S.D.O
and V.K.O) and  (in part) by
MEXT KAKENHI Grant-in-Aid for Scientific Research on Innovative Areas ``Cosmic
Acceleration''  (No. 15H05890) and the JSPS Grant-in-Aid for Scientific 
Research (C) \# 23540296 (S.N.).

\end{document}